# Interference-aided spectrum fitting method for accurately film thickness determination

Xingxing Liu(刘星星)[1,2,3], Shao-Wei Wang(王少伟)[1,2,*], Hui Xia(夏辉)[1], Xutao Zhang(张旭涛)[1,3], Ruonan Ji(冀若楠)[1,2,3], Tianxin Li(李天信)[1,2], Wei Lu(陆卫)[1,2]

[1]*National Laboratory for Infrared Physics, Shanghai Institute of Technical Physics, Chinese Academy of Sciences, Shanghai 200083, China*
[2]*Shanghai engineering research center of energy-saving coatings, Shanghai 200083, China*
[3]*University of Chinese Academy of Sciences, Beijing 100049, China*
\* Corresponding author: wangshw@mail.sitp.ac.cn

A new approach was proposed to accurately determine the thickness of film, especially for ultra-thin film, through spectrum fitting with the assistance of interference layer. The determination limit can reach even less than 1 nm. Its accuracy is far better than traditional methods. This determination method is verified by experiments and the determination limit is at least 3.5 nm compared with the results of AFM. Furthermore, double-interference-aided spectra fitting method is proposed to reduce the requirements of determination instruments, which allow one to determine the film thickness with a low precision common spectrometer and largely lower the cost. It is a very high precision determination method for on-site and in-situ applications, especially for ultra-thin films.

Nowadays, ultra-thin films have been widely used in optical or electronic devices. Gate dielectric films less than 5.0 nm thick are becoming common in MOS technology. [1] AZOs ultra-thin films about 10.0 nm play important roles in Low-E glasses. [2] Photoelectric properties of single and multi-layer MoS2 have been intensively studied for their value of vast potential applications. [3] Uniformity analysis of is helpful for large area nanoscale thin films produce. [4] It is especially important for optical thin films and will influence their optical property largely. Thus, ultra-thin film thickness determination, especially for measurement and controlling in the deposition process, is very important in industry and research fields. There are three main thickness determination methods for in-situ situation: quartz crystal oscillation method [5], ellipsometry method [6] and photometric method [7, 8].

Quartz crystal oscillation method is a good physical thickness monitoring method only when coating parameters are stable and small mass load. It is not a kind of accurate thin film monitoring method when process parameters appear to fluctuate. Furthermore, quartz crystal oscillation method cannot monitor the optical thickness and control the optical property of film. In order to monitor optical thickness accurately, ellipsometry method and photometric method were proposed. The former one is very sensitive with film thickness change which makes it as a good candidate for thickness measurement of ultra-thin film. [9] However, the ellipsometry system is complicated and can't use fiber to ensure the polarization property, which increase the difficulty for in-situ application and cost. Photometric spectrum is the main in-situ method for the detection and monitoring of film thickness up to now. [10]

Photometric method is a single wavelength monitoring method and outstanding when film thickness is comparable to λ/4 or thicker, where λ is the monitoring wavelength. For photometric spectrum (transmission or reflection spectra), spectral curves are easily calculated from the optical constants and thickness of the material through Transfer-Matrix Method (TMM) [11]. The optical constants and thickness can be fitted by some proper algorithms such as genetic algorithm [12] and annealing algorithm [13]. Predecessors have spent a lot of time in building the models [14] of optical constants and optimizing the algorithms. These efforts have improved the accuracy of thickness and index measurements largely. However, the calibration precision of thickness is not only affected by the fitting procedure, but also determined by the amount of information the spectrum contained and remarkably influenced by the precision of spectral obtaining process. When the thickness of film is very thin (generally smaller than 90 nm), the change of spectra with and without it is so small and no obvious characteristic spectral structure information exists leading to the difficulty of fitting and determination accurately. Furthermore, it is almost impossible to distinguish the film when its refractive index is close to that of substrate.Therefore, there is no suitable method for accurate thickness in-situ determination of films thinner than 90 nm.

We present an interference-aided spectrum fitting method to determine film thickness accurately with the help of interference layer, whose determination limit can be extended to even less than 1 nm. Furthermore, double-interference-aided spectra fitting simultaneously are introduced to lower the requirement for equipment's resolution.

All simulation and fitting processes in this work were carried out by CODE 3.27 of W. Theiss Hard and Software.

The structure of traditional fitting method is presented in Fig 1(a). The undetermined film is deposited on substrate directly in traditional method, and then its spectrum has been fitted to determine its thickness. This method is valid when the thickness of undetermined film is comparable to wavelength or thicker and the refractive index of undetermined film is quite different from the substrate. But it will lead to significant error when the film thickness is far less than the wavelength, because only small spectral intensity changes exist as shown in Fig. 1 (b), especially when the refractive index of film is close to that of the substrate. The refractive indexes of the substrate and undetermined film were set as 1.52 and 1.70 in the simulation. Meanwhile the thickness of the substrate was set as 1.0mm. Slightly change in intensity caused by undetermined films can be observed in traditional method from Fig 1(b). For example, the transmittance decreased only 0.2% to 91.7% at the wavelength of 400 nm when thickness of undetermined film is 6.0 nm. The transmittance is 91.5% when it is 10.0nm, which means the intensity change is only about 0.4%. Such a small transmittance change is lack of enough information to determinate the thickness accurately and easily influenced by the light source and external environment.

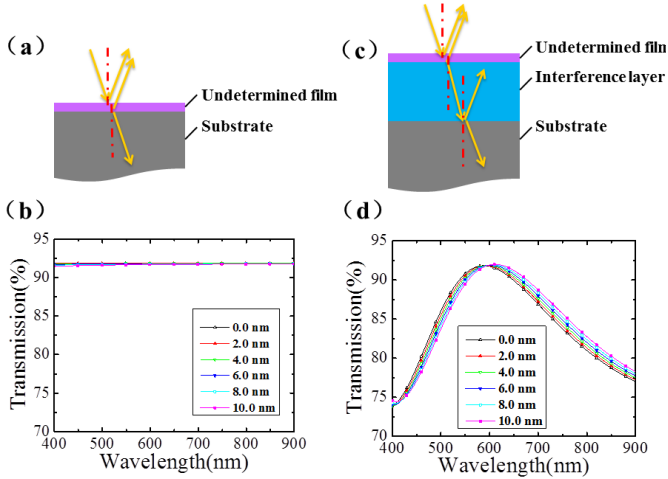

Fig.1 structure diagram of traditional method (a) and interference-aided spectrum fitting method (c); simulated transmission spectra of traditional method (b) and interference-aided spectrum fitting method (d) with different thickness undetermined films

To solve this problem, we propose an interference-aided spectrum fitting method by introducing an interference layer onto the substrate as Fig. 1(c) shows. The interference layer is thick enough to generate interference effect in transmission or reflection spectra when its refractive index is different from the substrate as shown in Fig. 1(d). The refractive index and thicknesses of the interference layer was set as 2.10 and 140.0 nm respectively in the simulation. Parameters of other layers were set as the same as the previous description. The transmittance increased from 73.9% to 74.6% at the wavelength of 400 nm when 10.0 nm undetermined film coated on, corresponding to intensity change of 0.9%, 2 times larger than that of traditional method. The peak position of 588.0nm resulted from the interference layer moves to 610.8 nm. The position change is 22.8nm and as high as 3.9%, almost 10 times to the intensity change of traditional method. Therefore, in addition to a more obvious intensity change, a much larger change of peak can be observed in the aid of interference layer. The peak shift of 4.6nm can still be obviously observed with only 2.0nm undetermined film as Fig. 1(d) shows. More importantly, the peak position almost determined only by the film property such as thickness and refractive index, and rarely affected by the measuring system and environment. Thus, the information contained and reliability of peak position measurement is much higher than that of intensity measurement, resulting in further improvement of determination accuracy with interference-aided spectrum fitting method.

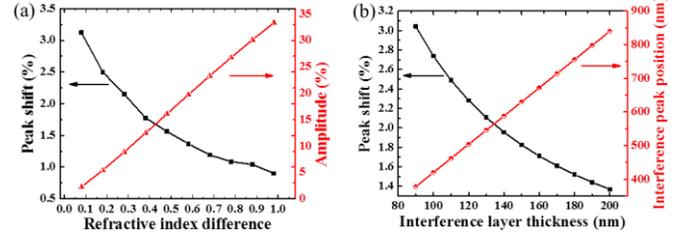

Fig.2 simulated peak shift and amplitude caused by undetermined film vs. refractive index difference between interference layer and substrate (a) ; simulated peak shift and peak position vs. thickness of interference layer (b). The refractive indexes and thicknesses of the substrate and undetermined film were set as 1.52 and 1.70, 1mm and 5.0 nm in this simulation, respectively.

To study the influences of thickness and refractive index of the interference layer, the relationships between peak shift, interference amplitude (the difference of transmittance between the interference peak and valley) and refractive index of the interference layer were simulated, as shown in Fig. 2(a). The thicknesses of undetermined film were set as 5.0 nm in this simulation. The interference peak shift results from undetermined film is reduced from 3.1% to 0.9% with the increase of refractive index difference between the substrate and interference layer from 0.08 to 0.98, when interference layer thickness is 200.0 nm as Fig. 2(a) shows. This decrease will reduce the determination accuracy of thickness. However, interference amplitude will increase with the increase of refractive index difference which is helpful for improving the determination accuracy. Therefore, appropriate refractive index of interference layer should be selected to balance the peak shift and interference amplitude. Although refractive index difference can change freely in theory, candidate materials in experiment are limited. $Al_2O_3$, $Si_3N_4$, $Nb_2O_5$, $Ta_2O_5$ and $TiO_2$ are some common weakly absorbing materials which can be used as interference layer. $Nb_2O_5$, $Ta_2O_5$ and $TiO_2$ are easily obtained high refractive index materials.

The relationships of peak shift and peak position with the thickness of interference layer were also studied as Fig. 2(b) shows, under the condition that the refractive index of interference layer is 2.10. There is no interference peak in visible band when the interference layer is thinner than 90 nm. With the increase of the interference layer thickness from 90 nm to 200 nm, the interference peak position is increased from 378 nm to 840 nm. In the meantime, the interference peak position shift caused by undetermined film is decreased from 3.0% to 1.4%. Furthermore, detector's sensitivity of spectrometer and appropriate interference peak position should be taken into account. CCD or CMOS detector is usually responding at the wave band from 400 nm to 1100 nm and the peak response is between 600 nm and 800nm [15]. Consequently, the interference layer thickness about 140nm is better for determination whose interference peak position is at the wavelength about 600 nm.

To study the method more practically, experimental level random noises were considered in both the traditional method and interference-aided spectrum fitting method. The SNR of spectrum was set to 1000:1 in experimental level. Meanwhile, the thickness of the interference layer was set as 140 nm in this simulation. Results of thickness determination of ultra-thin film are shown in Table I.

The determination limit is defined as the minimum thickness which can be determined with the error less than 20%. For $n_U$(undetermined film refractive index) =2.10, $d_U$ (thickness of undetermined film) determination using traditional method is valid over 10.0 nm, where the error is 11%. When undetermined film is 7.0 nm or thinner, traditional method is completely invalid. Thus, the determination limit of traditional method is about 10.0 nm in this case theoretically. But the determination limit of our method can reach 1.0 nm or even thinner when undetermined film refractive index is 2.10. Similarly, when undetermined film refractive index decreases to 1.70, the determination limit of our method is 2.0 nm compared to 15.0 nm of traditional method. Particularly, when the refractive index of undetermined film is the same as that of substrate, $n_U$ =1.52, ultra-thin film thickness is completely unable to be measured by using traditional method. However, it can still be determined by using interference-aided spectrum fitting method and the determination limit comes to 3.0 nm. Therefore, the determination limit of our method has been improved largely compared to traditional method. It is a good way for the determination of ultra-thin film thickness which fits for in-situ operation. All the improvement thanks to the help of introduced interference layer.

Table I Comparison of simulated thickness determination results between traditional and our methods with SNR of 1000:1.

| $d_U$ /nm | $d_U$/nm | | | | | |
|---|---|---|---|---|---|---|
| | $n_U$=2.10 | | $n_U$=1.70 | | $n_U$=1.52 | |
| | Traditional method (error) | Our method (error) | Traditional method (error) | Our method (error) | Traditional method (error) | Our method (error) |
| 15.0 | 14.9(-1%) | 15.0(0%) | 14.8(-1%) | 15.0(0%) | 59.3(293%) | 15.0(0%) |
| 10.0 | 11.1(11%) | 10.0(0%) | 15.1(51%) | 10.0(0%) | 59.0(490%) | 10.0(0%) |
| 7.0 | 8.7(24%) | 7.0(0%) | 15.2(117%) | 7.0(0%) | 139.4(1891%) | 7.0(0%) |
| 5.0 | 7.1(42%) | 5.0(0%) | 12.7(154%) | 5.0(0%) | 142.2(2744%) | 5.0(0%) |
| 3.0 | 7.3(143%) | 3.0(0%) | 8.9(197%) | 3.0(0%) | 132.2(4307%) | 3.0(0%) |
| 2.0 | 7.5(275%) | 1.9(-5%) | 9.0(350%) | 2.1(5%) | 203.7(10085%) | 1.4(-30%) |
| 1.0 | 7.6(660%) | 1.0(0%) | 9.1(810%) | 1.3(30%) | 254.2(25320%) | 1.5(50%) |

The method is also valid for reflection spectrum fitting in principle. However, since reflection spectrum is an indirect measurement and will be affected by many factors such as incident angle and reference mirror, its accuracy and reliability is lower than transmission ones. Therefore, it is used only in case of where the transmission one is invalid or very difficult to be realized, for instance of opaque substrates or system where transmission signal can't be obtained.

Based on above theoretical analysis, interference layer-aided spectrum fitting method is a prospective way to determinate the thickness of thin film precisely with limit less than 1 nm. The determination procedure of thickness can be conducted as follows: Transmission spectrum of the interference layer coated substrate is carried out by a spectrometer as standard reference. The thickness of undetermined film can be obtained by fitting the transmission spectra before and after coating the undetermined film onto the substrate with interference layer.

To prove the validity of our method, a batch of $Nb_2O_5$ and $TiO_2$ interference layers were deposited by Leybold ARES1110 High Vacuum Coating System as standard pieces. The process is similar to ref. 12. $SiO_2$ with different thickness, as undetermined films, were deposited on sub/$Nb_2O_5$, and Si (0 0 1) substrate simultaneously to form steps of $SiO_2$ which can be measured by AFM as comparison.

Transmission spectra of samples in visible and NIR band were measured by Lambda 950 spectrometer of PerkinElmer. Thickness of $SiO_2$ was measured by DI multimode scanning probe microscopy with IV controller.

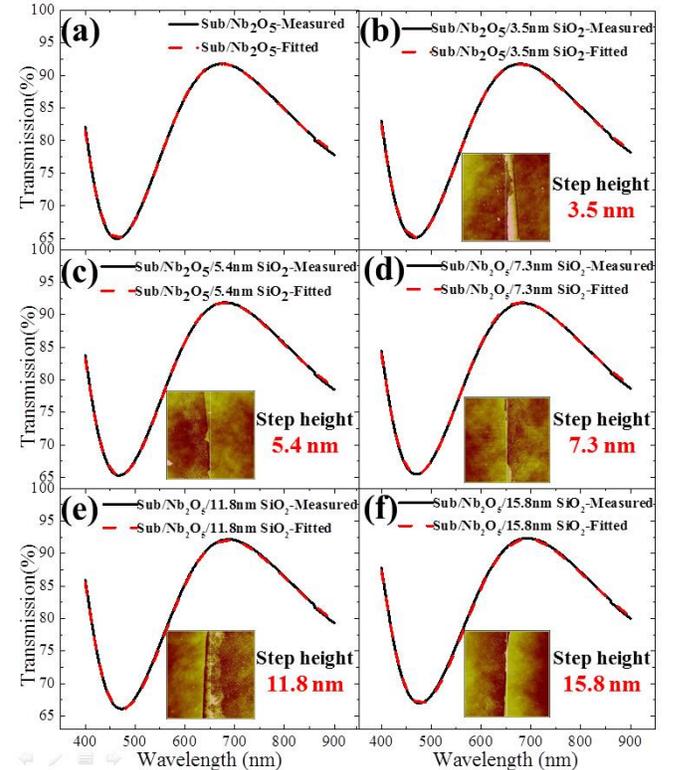

Fig.3 measured (taken by lambda 950) and fitted transmission spectra of interference layer $Nb_2O_5$ deposited on K9 glass (a); measured (taken by lambda 950), fitted transmission spectra and AFM results of 3.5 nm, 5.4 nm, 7.3 nm, 11.8 nm and 15.8 nm ultra-thin films on interference layer (b), (c), (d), (e) and (f).

Cauchy model with exponential absorption is useful for dielectric materials far from the absorption bands [16]. The formulas are:

$$n(\lambda) = n_0 + \frac{n_1}{\lambda^2} + \frac{n_2}{\lambda^4}, \quad k(\lambda) = k_0 \exp(\frac{k_1}{\lambda})$$

The spectra of $Nb_2O_5$ and $SiO_2$ films on these samples were fitted using this material model to describe their optical constants. By fitting the spectrum of each sample, thickness and refractive index of undetermined film will be carried out directly. The measured (taken by lambda 950) and fitted transmission spectra of interference layer $Nb_2O_5$ deposited on K9 glass are presented in Fig. 3(a). The thickness of $Nb_2O_5$ is 148.9 nm. Its refractive index is about 2.28 at the wavelength of 600 nm. The spectra of samples with different thickness of undetermined SiO$_2$ films were all fitted perfectly as Fig. 3(b) ~ (f) shows. Undetermined film thickness is 3.1 nm, 5.6 nm, 7.8 nm, 11.1 nm and 16.5 nm by fitting, respectively. These results agree very well to AFM results of 3.5 nm, 5.4 nm, 7.3 nm, 11.8 nm and 15.8 nm, as show in the insets of Fig. 3(b) ~ (f). Thus, thicknesses of undetermined films can be determined by interference-aided spectrum fitting method rapidly and accurately, even for only 3.5 nm. The introducing of interference layer overcomes the shortcomings of traditional fitting methods through transmission or reflection spectra.

The interference-aided spectrum fitting method can determine the thickness precisely especially for ultra-thin films, but it needs high precision spectrometer to ensure the measured results because it is still strongly dependent on intensity value. It is useful in ex-situ detection through high precision spectrometer. But it is expensive and most of online spectroscopy systems don't have so high precision. In order to extend this method for in-situ operation, requirement for precision of spectrometer should be reduced.

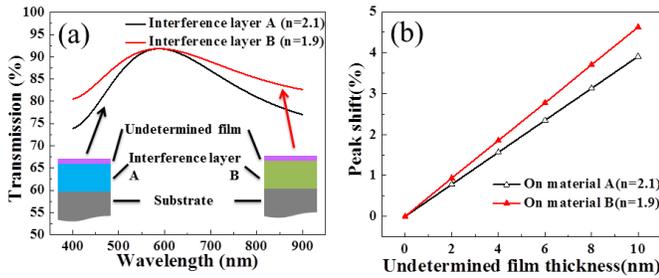

Fig.4 transmission spectra of interference layers n=1.9 and 2.1 and structure diagram of double-interference-aided spectra fitting method (a); peak shift vs. thickness of undetermined films on interference layers n=1.9 and 2.1 (b).

The above studies show that the peak shifts are different when refractive indexes of the interference layers are different, as shown in Fig. 4(a) and (b). Thus, more information can be obtained if using two different interference layers simultaneously. In this case, there are only two unknown quantities $n_U$ and $d_U$ with two independent spectra, as shown in Fig. 4(a). They can carry out two independent equations and the $n_U$ and $d_U$ can be solved separately in principle. When fitted with the actual spectra generated by two different kinds of interference layers simultaneously, the information quantity can be further increased and the requirement for spectrometer should be able to be reduced to a certain extent.

A double-interference-aided spectra fitting method has been introduced based on the above mechanism analysis. It can be conducted similar to that of single-interference-aided spectrum fitting method. The only difference is two substrates with different interference layers should be put into coating system to deposit the same ultra-thin film simultaneously. Transmission spectra (or reflection spectra) of these samples can be carried out by a spectrometer in turn. Then the spectra are fitted simultaneously to get the thickness of undetermined film.

To validate this approach, Nb$_2$O$_5$ and TiO$_2$ are chosen as two interference layers. Then, they were put into the coating system together to deposit the undetermined SiO$_2$ film simultaneously. High precision spectrometer of lambda 950 and low precision fiber micro spectrometer of IdeaOpticals PG 2000 are compared in this work. Table II presents the results of thickness determination by different methods. When fitted with spectra measured by low precision PG2000, it is unable to determine the film thickness even for 15.8 nm with error larger than 20%. When using double-interference-aided method with PG 2000, it can give a reasonable result of 15.5 nm, which thanks to the help of another interference layer. It is similar for other thicknesses even for 3.5 nm which can give the result of 3.2 nm. Therefore, the determination limit of double-interference-aided method can also reach to 3.5 nm with low precision measuring system. Therefore, double-interference-aided method can effectively lower the requirement for measuring instrument and reduce the cost largely. The low precision fiber micro spectrometer is often used for in-situ detection. Therefore, double-interference-aided spectra fitting method is promising for low cost on-line monitoring.

For pratical application of the method, the substrate with interference layer can be fabricated in mass production and easily used as standard reference just like quartz monitor crystal. One just need to measure the spectra before and after depositing the undetermined film and fit them.

Table II the results of ultra-thin film thickness determined with different methods experimentally

| Method | Thickness of SiO$_2$ (nm) | | | | | |
|---|---|---|---|---|---|---|
| AFM | 3.5 | 5.4 | 7.3 | 11.8 | 12.9 | 15.8 |
| Lambda 950 with interference-aided method | 3.1 | 5.6 | 7.8 | 11.1 | 13.5 | 16.5 |
| Fiber spectrometer with interference-aided method | 1.3 | 1.4 | 3.1 | 5.8 | 10.1 | 12.2 |
| Lambda 950 with double-interference-aided method | 3.4 | 5.5 | 7.6 | 11.3 | 13.1 | 15.9 |
| Fiber spectrometer with double-interference-aided method | 3.2 | 5.1 | 6.8 | 10.9 | 13.6 | 15.5 |

Based on the improvement from interference-aided spectrum fitting method to double-interference-aided spectra fitting method, three or more different interference layers may be used to meet different measuring demands. Furthermore, a determination chip that integrated multi interference layers should be helpful in thickness determination for any unknown films, which can provide more information for analysis.

In summary, an interference-aided spectrum fitting method is proposed to determine film thickness precisely. Interference layer is introduced to form interference peak in the transmission (or reflection) spectrum. It can result in obvious interference peak shift and enlarge intensity change of spectrum several times which is helpful to improve the thickness determination precision and limit of thin film. The determination limit can reach less than 1nm theoretically. It is verified by experiments and the determination limit is at least 3.5 nm. Furthermore, double-interference-aided fitting method has also been developed to lower the requirement for measuring system and reduce the cost largely. It provides a prospective way for very high precision on-site and in-situ thickness determination, especially for ultra-thin films.

This work was partially supported by the Shanghai Science and Technology Foundations (13JC1405902, 15DZ2282100, 16DZ2290600), Youth Innovation Promotion Association CAS (2012189) and National Natural Science Foundation of China (61223006, 61376053).